\documentclass[preprint]{revtex4}
\pdfoutput=1
\usepackage[dvips]{graphicx}
\usepackage{graphicx}
\usepackage{amsmath}
\usepackage{amssymb}
\usepackage{amsfonts}
\usepackage{upgreek}
\usepackage{txfonts}
\usepackage{color}
\usepackage{slashed}
\usepackage{latexsym}
\usepackage{amsmath}
\usepackage{bm}
\usepackage{soul}
\newcommand{\beq}{\begin{equation}}
\newcommand{\eeq}{\end{equation}}
\newcommand{\ba}{\begin{array}{ccc}}
\newcommand{\ea}{\end{array}}
\newcommand{\nn}{\nonumber \\}

\newcommand{\bk}{{\bm k}}
\newcommand{\bp}{{\bm p}}
\newcommand{\bq}{{\bm q}}
\newcommand{\bl}{{\bm l}}
\def\bea{\begin{eqnarray}}
\def\eea{\end{eqnarray}}
\usepackage{hyperref}
\usepackage{xcolor}
\definecolor{dark-red}{rgb}{0.4,0.15,0.15}
\definecolor{dark-blue}{rgb}{0.15,0.15,0.4}
\definecolor{medium-blue}{rgb}{0,0,0.5}
\hypersetup{ colorlinks, linkcolor={dark-blue},citecolor={dark-blue}, urlcolor={medium-blue} }

\begin{document}
\setstcolor{red}

\title{
Stress tensor and current correlators of interacting conformal field theories in 2+1 dimensions: 
Fermionic Dirac matter coupled to $U(1)$ gauge field
}

\author{Yejin Huh$^{1,2}$}
\email{yhuh@physics.utoronto.ca}
\author{Philipp Strack$^{1,3}$}
\email{pstrack@physics.harvard.edu}
\homepage{http://www.thp.uni-koeln.de/~strack/}

\affiliation{$^{1}$Department of Physics, Harvard University, Cambridge MA 02138}
\affiliation{$^{2}$Department of Physics, University of Toronto, Ontario M5S 1A7, Canada}
\affiliation{$^{3}$Institut f\"ur Theoretische Physik, Universit\"at zu K\"oln, D-50937 Cologne, Germany}

\date{\today}

\begin{abstract}
We compute the central charge $C_T$ and universal conductivity $C_J$ of $N_F$ fermions coupled 
to a $U(1)$ gauge field 
up to next-to-leading order in the $1/N_F$ expansion. 
We discuss implications of these precision computations as a diagnostic for response and 
entanglement properties of interacting conformal field 
theories for strongly correlated 
condensed matter phases and conformal quantum electrodynamics in $2+1$ dimensions.
\end{abstract}

\maketitle

\tableofcontents

\section{Introduction}
A variety of strongly correlated electron systems 
at quantum critical points or phases in two spatial dimensions are believed 
to be described by (interacting) conformal field theories in 2+1 dimensions (CFT$_3$'s). 
The workhorse is the Wilson-Fisher CFT$_3$, also known as the $O(N)$-model of a 
real-valued vector field with $N$ components \cite{wilsonfisher71,abe73,petkou95}, which 
describes, among other things, the Ising model for $N=1$ \cite{showk12,showk14}, 
superfluid-to-insulator transitions for $N=2$ \cite{cha91,fazio96}, and quantum magnetic transitions for $N=3$ 
\cite{nelson89,chubukov94}.
Especially intriguing are gauge theoretical descriptions of condensed matter systems
(e.g.: \cite{kaul08} and references therein for an overview) such as
 of quantum Hall systems 
(e.g.: \cite{chen92,chen93} and references therein),
fractionalized magnets and deconfined critical points in strongly correlated Mott insulators 
\cite{senthil04,sandvik07,huhprl13}, and effective theories for the cuprates 
\cite{rantner02,franz02,franz03,kaul07}. There, the relevant dynamics is often provided 
by emergent or effective degrees of freedom not necessarily present in the bare Hamiltonian.
These conformal phases of quantum matter in 2+1 dimensions provide a unique interpolation 
between the better understood CFT's in 1+1 dimensions \cite{cardy08} and  
much studied gauge theories for high energy vacua in 3+1 dimensions 
\cite{polyakov87,coleman88}.

A common feature of CFT$_3$'s is the absence of quasi-particles and for 
condensed matter systems it is 
of particular interest to understand
response properties of interacting CFT$_3$'s to externally applied perturbations such as electromagnetic 
fields or mechanical forces {\it without invoking a quasi-particle picture}.

\subsection{Model: $N_F$ Dirac fermions coupled to $U(1)$ gauge field}

In this paper, we consider $N_F$ Dirac fermions minimally coupled to a $U(1)$ gauge field. 
This theory arises in a variety of condensed matter contexts \cite{kaul08,chen93,rantner02,franz02, kaul07}.
The Euclidean action,
\begin{align}
\mathcal{S} = \int d^2 r d \tau \bar{\psi}_\alpha \left[ i \gamma^{\mu}
 \left( \partial_\mu - i \frac{A_\mu}{\sqrt{N_F}} \right ) \right] \psi_\alpha + ...\;,
\label{eq:bare_action}
\end{align}
contains Grassmannian two-component fermion fields $\bar{\psi}_\alpha$ and $\psi_\alpha$, where $\alpha$ is the fermion flavor index, and $\mu$ is the spatial and (imaginary) temporal index in 
$2+1$ dimensions. 
Repeated indices are summed over. $\gamma^{\mu}$'s are the Dirac matrices that satisfy 
$\{ \gamma^{\mu} ,\gamma^{\nu}\} = 2 \delta^{\mu\nu}$. We use the same conventions as Kaul 
and Sachdev for their fermion sector \cite{kaul08}. The dots stand for additional terms 
which may play a role in the UV and away from the conformally invariant fixed-point considered 
in this paper.

The gauge field $A_\mu$, a conventional spin-1 boson often dubbed as ``emergent photon'' 
in the condensed matter context, 
ensures fulfillment of a local $U(1)$ gauge symmetry at every point  $(\tau,\mathbf{r})$ in (Euclidean) 
space-time. A potential, bare Maxwell term $\frac{1}{2 e^2} F_{\mu\nu} F^{\mu\nu}$ 
is not written in Eq.~(\ref{eq:bare_action}) and is unimportant for the universal constants at the infrared 
fixed point of interest in this paper \cite{klebanov12}.
The gauge field gets dynamical by integrating the fermion fields in the large $N_F$ limit. 
In Landau gauge, the gauge field propagator at $N_F \rightarrow \infty$ is purely transverse and takes the
characteristic overdamped form (with $p=|\bp|$)
\begin{align} 
D^{(0)}_{\mu\nu} (p)= \frac{16}{p} \left( \delta_{\mu\nu} - \frac{p_\mu p_\nu}{p^2}\right)\;.
\label{eq:bare_gauge}
\end{align} 
Model Eq.~(\ref{eq:bare_action}) with a bare Maxwell term is also known as 
QED$_3$ and flows to strong coupling in the infrared and shares its propensity to form
fermion bound states ``mesons'' with QCD in 3+1 dimensions \cite{appelquist86,appelquist88,nash89}.
Deforming QED$_3$ toward graphene-type models with instantaneous Coulomb interactions 
are also interesting \cite{son07,juricic10,herbut13, kotikov14, barnes14}. 
It is believed that for sufficiently large $N_F$, Eq.~(\ref{eq:bare_action}) flows to a strongly coupled 
conformal phase in the infrared, preserving scale invariance \cite{braun14} (and references therein).
%This is the regime of interest in the present paper.

\subsection{Key results: central charge $C_T$ and $C_J$ up to next-to-leading order in $1/N_F$}

The main result of this paper is an explicit formula and numerical value of the central charge $C_T$ 
of Eq.~(\ref{eq:bare_action}), defined below as the universal constant appearing 
in the stress tensor correlator at the interacting conformal fixed point,
up to next-to-leading order in the $1/N_F$ expansion:
\begin{align}
\frac{C_T}{N_F} = &\frac{1}{256} \left(1 + \frac{1}{N_F} \left( \tilde{C_T}^{(1)} + \frac{104}{15\pi^2}\right)\right)\;
\nonumber\\
=&
\frac{1}{256}\left(1 + \frac{0.28701185900024704065}{N_F}\right)\;.
\label{eq:CT_final}
\end{align}
$\tilde{C_T}^{(1)}$ comes from one out of nine Feynman graphs in momentum space computed below in Fig.~(\ref{fig:diagsTT}) 
\begin{align}
\tilde{C_T}^{(1)} & =-\frac{4}{45 \pi^2} \Bigg( 180 \text{Li}_2\left(3-2 \sqrt{2}\right)-720
   \text{Li}_2\left(-1+\sqrt{2}\right)-398+90 \pi ^2+45 \log ^2\left(3-2
   \sqrt{2}\right)
      \nonumber\\
& ~~~~~~~~~~~~~~~~
   +1146 \sqrt{2} \log \left(3-2 \sqrt{2}\right)+  
  12 \left(191
   \sqrt{2}+15 \log \left(3-2 \sqrt{2}\right)\right) \sinh ^{-1}(1) \Bigg)\nn
& = ~ -0.41548168091996150803
   \;,
   \label{eq:ct1}
\end{align}
where Li$_{n}(z) = \sum_{k=1}^{\infty} \frac{z^k}{k^n}$ is the polylogarithm or Jonqui\'ere's function 
for $n=2$. The sum of other eight diagrams evaluate to the remaining term in the 
innermost bracket, $\frac{104}{15 \pi^2}$, in the first line of Eq.~(\ref{eq:CT_final}). We observe from Eq.~(\ref{eq:CT_final}) that $1/N_F$ corrections 
to the $N_F\rightarrow \infty$ value are $\approx 15\%$  when $N_F = 2$. Even larger $1/N$ corrections 
were observed (for current correlators) in the $CP^{N-1}$ model and attributed in particular 
to vertices directly involving the gauge field \cite{huh13_long}.

It is interesting to note that here in Eq.~(\ref{eq:CT_final}) 
the $1/N_F$ corrections are positive whereas in certain theories with bosonic field content \cite{huhprl13,huh13_long, petkou94,petkou95,petkou96} the $1/N_B$ corrections to $C_T$ as well as $C_J$ (see below) 
typically turn out to be negative. Given this information, the sign of the correction could be attributed 
to the quantum statistics of the charged fields but further analysis (see also conclusions for an outlook ondual 
Chern-Simons + matter theories) and potentially higher-order computations are needed to uncover further 
the structure of these corrections.

It is hard to overestimate the fundamental importance of the central charge in conformal 
field theory with applications ranging from thermodynamics, quantum critical transport, to 
quantum information theory \cite{cardy10}. An interesting recent application 
are explicit formulae for the R\'enyi entropy for $d$-dimensional flat space CFT's and 
we quote here the formula from Perlmutter \cite{perlmutter14}:
\begin{align}
S'_{q=1}= - {\rm{Vol}}\left(\mathbb{H}^{d-1}\right) \frac{\pi^{d/2+1} \Gamma(d/2) (d-1)}{(d+1)!} C_T\;.
\end{align}
The prime denotes a derivative with respect to $q$ of the R\'enyi entropy 
$S_q = \frac{1}{1-q} \log {\rm Tr} \left[\rho^q\right]$, $\rho$ a reduced density matrix, 
and $\mathbb{H}^{d-1}$ the hyperboloid entangling surface. Moreover, precision values of $C_T$ 
may be useful for conformal bootstrap approaches for the 3D-Ising and other models \cite{showk12} 
as well as serving as a benchmark for numerical simulations of frustrated quantum 
magnets \cite{kaul13}.

In the present paper, we compute $C_T$ by direct evaluation of Feynman graphs 
in momentum space fulfilling and using the relation
\cite{cardy87,chowdhury13},
\begin{align}
\langle T_{\mu\nu}(-p)T_{\lambda\rho}(p) \rangle = C_T |\bp|^3 \Bigg(&  \delta_{\mu\lambda}\delta_{\nu\rho} + \delta_{\mu\rho}\delta_{\nu\lambda} - \delta_{\mu\nu}\delta_{\lambda\rho} + \delta_{\mu\nu}\frac{p_\lambda p_\rho }{p^2} 
+ \delta_{\lambda\rho}\frac{p_\mu p_\nu }{p^2} 
- \delta_{\mu\lambda}\frac{p_\nu p_\rho}{p^2} 
- \delta_{\nu\lambda}\frac{p_\mu p_\rho }{p^2}
\nonumber\\
&
 - \delta_{\mu\rho}\frac{p_\nu p_\lambda }{p^2} - \delta_{\nu\rho}\frac{p_\mu p_\lambda }{p^2} + \frac{p_\mu p_\nu p_\lambda p_\rho}{p^4}       \Bigg)\;
 \label{eq:TTgeneral}
\end{align}
generalizing our recently developed technology \cite{huhprl13,huh13_long} to Dirac fermions and 
contractions over stress tensor vertices. We discuss this further in Sec.~\ref{sec:TT}.

Computations of stress tensor correlators in interacting CFT's (at least 
without an excessive amount of symmetry such as supersymmetries) 
in effective dimensionality greater than 2  
are extremely scarce and we are not aware of a previous computation of $C_T$ for 
Eq.~(\ref{eq:bare_action}) in 2+1 dimensions. We quote here related works across the quantum field theory 
universe we are aware of to date:
two papers by Hathrell using loop expansions from 1982, one on scalar fields up to 
5-loops \cite{hathrell_scalar} and one on QED up to 3 loops \cite{hathrell_qed},
a two-loop analysis for general gauge theories coupled to fermions and scalars 
in curved space by Jack and Osborn in 1984 and 1985 \cite{jack84,jack85}, an $\epsilon$-expansion  
around four dimensions for scalar and gauge theories by Cappelli, Friedan and 
LaTorre in 1991 \cite{cappelli91}, 
and a series of papers on the $O(N)$ vector model 
from 1994-1996 by Petkou and Osborn \cite{petkou94,petkou95,petkou96}, 
and a three-loop OPE computation in massless QCD by Zoller and Chetyrkin in 2012 \cite{zoller12}.

For essentially free field theories, stress tensor amplitudes \cite{maldacena11,chowdhury13} and 
R\'enyi entropies \cite{klebanov12} have also been computed.
(Multi-point) correlators of the stress tensor are also instrumental for the relation between 
scale and conformal invariance (e.g.: \cite{dymarsky13,bzowski14}). It would be interesting 
to consider generalizations of Eq.~(\ref{eq:bare_action}) with conformally invariant UV fixed points 
to be able to compare $C_T^{IR}$ and $C_T^{UV}$ for a given number of flavors 
in the context of generalized c-theorems for CFT's  in general dimensions \cite{myers11,klebanov11,appelquist99}.
It is known that QED$_3$, including a Maxwell term $\frac{1}{2e^2} F^2$, 
flows toward a weakly interacting UV fixed-point. Against this backdrop, an assessment of the full 
conformal symmetry (free photons are not necessarily conformally invariant in the UV), and a systematic 
investigation of possible UV fixed points and their relevant operators is an interesting extension 
of our work.

%\subsection{Key result: $C_J$ next-to-leading order in $1/N_F$}

The second result of this paper is an (somewhat simpler) computation 
of the universal constant $C_J$ of the two-point correlator of the conserved flavor 
current of Eq.~(\ref{eq:bare_action}):
\begin{align} 
J^\ell_\mu = \bar{\psi}_\alpha T^\ell_{\alpha\beta} \gamma_\mu \psi_\beta\;,
\end{align} 
where $T^\ell$'s are generators of the SU$(N_F)$ group normalized to satisfy $\text{Tr}(T^\ell T^m) = \delta^{\ell m}$. 
As the stress tensor $T_{\mu\nu}$, this flavor current is conserved and its two-point correlator depends on one 
universal constant $C_J$
\begin{align} 
\langle J^\ell_\mu(-p) J^m_\nu(p) \rangle = -C_J |\bp| \left( \delta_{\mu\nu} - \frac{p_\mu p_\nu}{p^2} \right) \delta^{\ell m} \;.
\label{eq:JJform}
\end{align}
For single fermion QED$_3$, $C_J$ describes the universal electrical conductivity in the collisionless regime $\omega \gg T$, 
with $T$ being the temperature. Depending on the physical context, however, it may also be related to magnetic or other 
response functions \cite{rantner02}. Our result for $C_J$ to next-to-leading order in $1/N_F$ is (derived in Sec.~\ref{sec:JJ})
\begin{align} 
C_J=&
\frac{1}{16} \left( 1 + \frac{1}{N_F} \left(\tilde{C_J}^{(1)} - \frac{40}{9\pi^2}\right)\right)
\nonumber\\
=&\frac{1}{16} \left( 1 + \frac{1}{N_F} 0.14291062004225554348\right)\;
\label{eq:value_cj}
\end{align}
with the analytical expression corresponding to one of the graphs being
\begin{align}
\tilde{C_J}^{(1)} = & -\frac{4}{3 \pi^2}\Bigg( -34 + 6\pi^2 + \sinh^{-1}(1) \left( 52 \sqrt{2} + 6 \log \left( 17-12\sqrt{2} \right)  \right) +26 \sqrt{2} \log \left(3-2\sqrt{2}\right) \nn
& +3 \log ^2\left(3-2 \sqrt{2}\right)+ 24 \text{Li}_2\left(1-\sqrt{2}\right)-24 \text{Li}_2\left(-1+\sqrt{2}\right) \Bigg)\nn
= & ~0.59322699178597897212\;.
   \label{eq:JJ1}
\end{align}
As for $C_T$, we again find the $1/N_F$ corrections to $C_J$ to be positive in contrast to the bosonic 
field theories analyzed in Ref.~\onlinecite{huhprl13,huh13_long}. Our numerical value of the correction is seemingly in disagreement with the value computed in the Appendix of Ref.~\onlinecite{chen93} and 
we compare to their value in detail in Sec.~\ref{sec:JJ}. As a (positive) cross-check, we have repeated a different calculation of the (non-conserved) staggered spin susceptibility in the Appendix 
of Rantner and Wen \cite{rantner02} using our approach and found the same logarithmically divergent coefficients.

Note that Eq.~(\ref{eq:bare_action}) has a further conserved ``topological'' current related to the curl of the gauge field 
\cite{chowdhury13} but we do not consider it further here.

\subsection{Organization of paper}

The remainder of the paper is organized as follows: in Sec.~\ref{sec:JJ}, we define the Feynman rules for Eq.~(\ref{eq:bare_action}) 
and the current vertex, and evaluate the 3 graphs renormalizing the current-current correlator.
In Sec.~\ref{sec:tensoria}, we briefly recapitulate the main elements of the Tensoria technology for the momentum integrals.
In Sec.~\ref{sec:TT}, we define the stress tensor vertex and evaluate the 9 graphs renormalizing the stress tensor 
correlator. In the conclusions, we summarize and point toward potential future directions where our technology could be applied to.

%\newpage
\section{Flavor current correlator $\langle J J \rangle$}
\label{sec:JJ}

In this section, we compute the SU$(N_F)$ flavor current-current correlator 
and compare it to the previous computation also using the $1/N_F$ expansion 
that we are aware of \cite{chen93}. We begin by stating the Feynman rules, 
compute the leading $N_F \rightarrow \infty$ graph in some detail, and then the more 
complicated self-energy and vertex corrections at order $1/N_F$. We will separate the 
contributions into longitudinal and transverse projections and show that all longitudinal 
and logarithmically singular corrections mutually cancel as they should for a conserved, 
transverse quantity.
%We will also recapitulate the technology that we use 
%\cite{huhprl13,huh13_long} to evaluate the tensor-valued integrals in momentum 
%space using Davydychev permutation relations \cite{davydychev91,davydychev92,bzowski12}.

\subsection{Feynman rules and graphs in momentum space}
\begin{figure}[t]
\includegraphics[width=120mm]{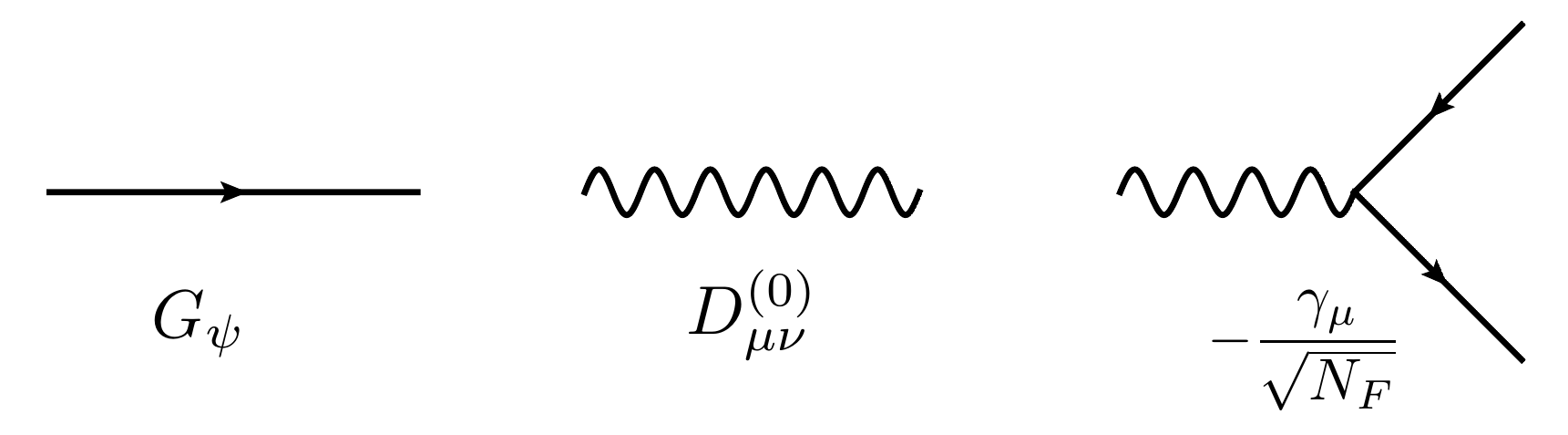}
\caption{Feyman rules for $N_F$ Dirac fermions coupled to $U(1)$ gauge field in Eq.~(\ref{eq:bare_action}).}
\label{fig:feyn1}
\end{figure}

\begin{figure}[t]
\includegraphics[width=40mm]{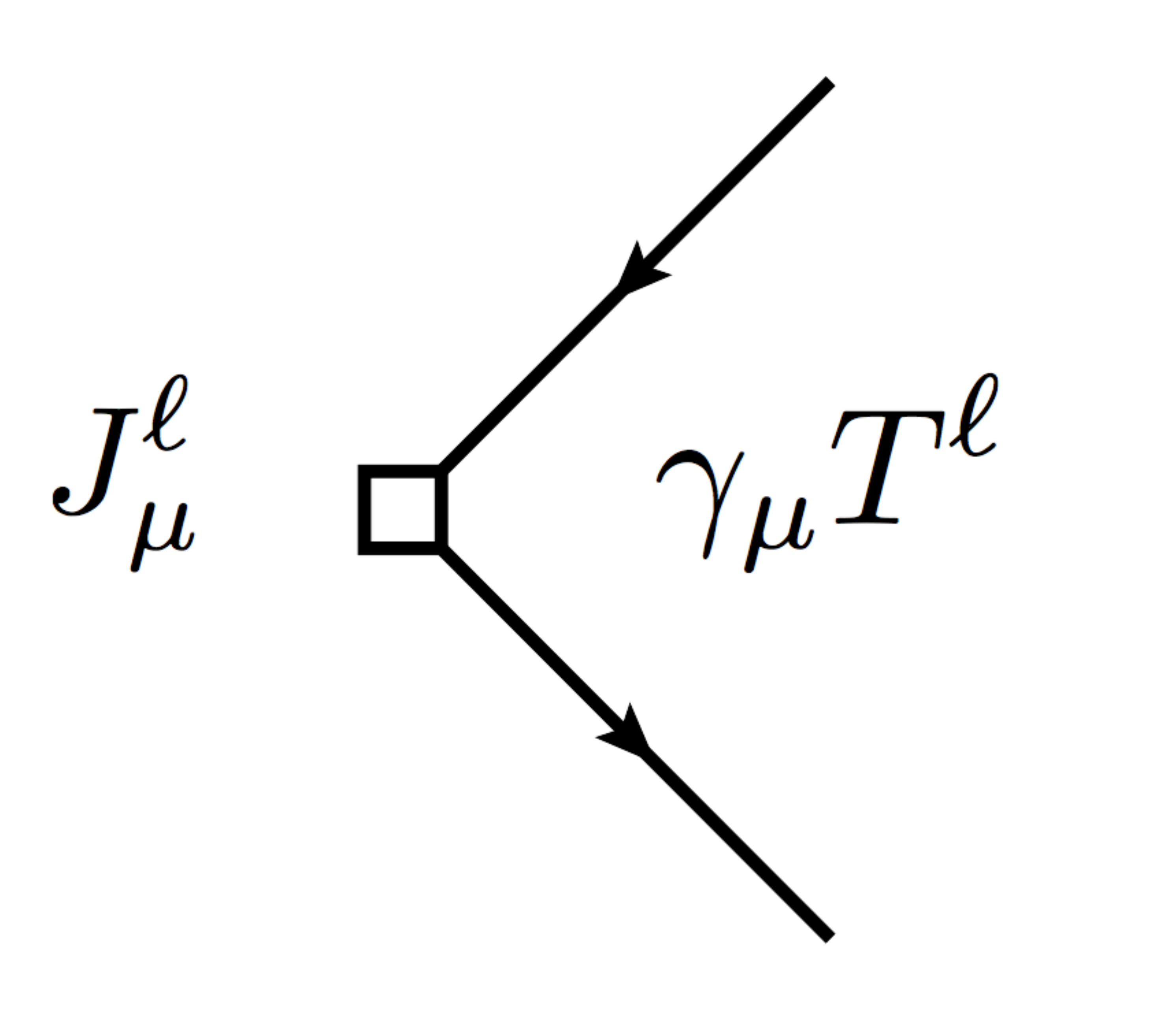}
\caption{Feyman rule for the current vertex. $T^{\ell}$ is a generator of the SU$(N_F)$.}
\label{fig:current}
\end{figure}

The Feynman rules for $N_F$ Dirac fermions coupled to $U(1)$ gauge field in Eq.~(\ref{eq:bare_action}) 
contain the relativistic fermion propagator
\begin{align}
G_{\psi}(k) = \frac{ k_a \gamma_a}{k^2}\;,
\end{align}
the gauge field propagator in Eq.~(\ref{eq:bare_gauge}), and the photon-fermion 
vertex drawn in Fig.~\ref{fig:feyn1}.
The current vertex in Fig.~\ref{fig:current} involves one generator of the SU$(N_F)$ but the traces over them 
in the actual diagrams are innocuous and just give $\delta$-functions in the flavor indices.

\begin{figure}[b]
\includegraphics[width=140mm]{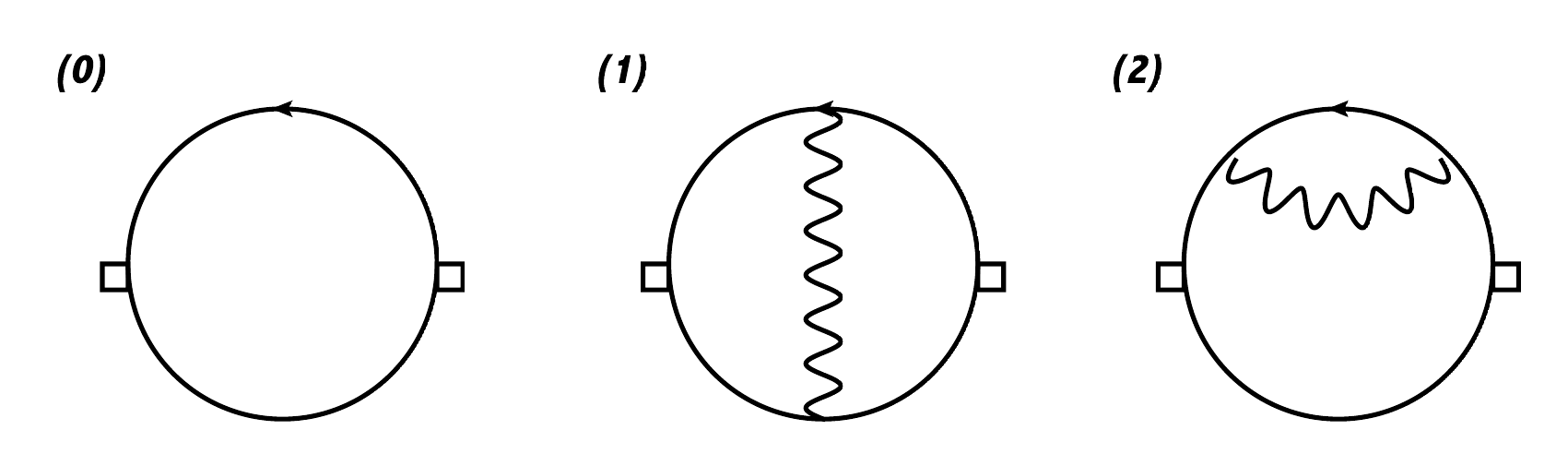}
\caption{Feyman diagrams contributing to the current current correlator to order $1/N_F$. Diagram (0) is the leading order contribution and 
the only one that survives the $N_F \rightarrow \infty$ limit. Diagram (1) is the vertex correction, diagram (2) the self-energy correction that comes with a factor of $a_2 = 2$. }
\label{fig:diagsJJ}
\end{figure}

Using the Feynman rules explained above, Fig.~\ref{fig:diagsJJ} exhibits the 3 contractions 
to the current correlator to order $1/N_F$. Each of the expressions in Eq.~\ref{eq:JJ} contain a minus sign due to the trace over fermions, a (trivial) trace over flavor indices, a trace over the Dirac matrices, 
and one (1-loop graph) or two (the two 2-loop graphs) $2+1$ dimensional momentum integrals $\int_{\bk} \equiv \int \frac{ d^3 k}{8 \pi^3}$.
We get:
\begin{align} 
\label{eq:JJ}
J_{\mu\nu}^{\ell m}(p)^{(0)} &= -\text{Tr}\left[ \int_\bk \gamma_\nu T^m \frac{k_a \gamma_a }{k^2} \gamma_\mu T^\ell \frac{(\bp+\bk)_b \gamma_b}{(\bp+\bk)^2}\right] 
%& a_0 = 1\nn
\\
J_{\mu\nu}^{\ell m}(p)^{(1)} &= -\text{Tr}\left[ \int_{\bk,\bq} \gamma_\nu T^m \frac{(\bk+\bp)_a \gamma_a}{(\bk+\bp)^2} \frac{\gamma_\lambda}{\sqrt{N_F}}  \frac{(\bk+\bp+\bq)_b \gamma_b}{(\bk+\bp+\bq)^2} \gamma_\mu T^\ell \frac{(\bk+\bq)_c \gamma_c}
{(\bk+\bq)^2} \frac{\gamma_\rho}{\sqrt{N_F}} \frac{k_d \gamma_d}{k^2}
\frac{16}{q}\left( \delta_{\lambda\rho} - \frac{q_\lambda q _\rho}{q^2}\right)      \right]  
%& a_1 = 1 \nn
%
\nonumber\\
J_{\mu\nu}^{\ell m}(p)^{(2)} &= -\text{Tr}\left[ \int_{\bk,\bq} \gamma_\nu T^m \frac{(\bk+\bp)_a \gamma_a}{(\bk+\bp)^2} \gamma_\mu T^\ell \frac{k_b \gamma_b}{k^2} \frac{\gamma_\rho}{\sqrt{N_F}} \frac{(\bk+\bq)_c \gamma_c}{(\bk+\bq)^2} \frac{\gamma_\lambda}{\sqrt{N_F}} \frac{k_d \gamma_d}{k^2} 
\frac{16}{q} \left( \delta_{\lambda\rho} - \frac{q_\lambda q _\rho}{q^2}\right)      \right] \;.
%J_{\mu\nu}^{\ell m}(p)^{(2)} &= -\text{Tr}\left[ \int_{\bk,\bq} \gamma_\nu T^m \frac{(k+p)_a \gamma_a}{(k+p)^2} \gamma_\mu T^\ell \frac{k_b \gamma_b}{k^2} \frac{\gamma_\rho}{\sqrt{N}} \left(  \frac{(k+q)_c }{(k+q)^2} - \frac{q_c}{q^2}\right) \gamma_c\frac{\gamma_\lambda}{\sqrt{N}} \frac{k_d \gamma_d}{k^2} 
%\frac{16}{q} \left( \delta_{\lambda\rho} - \frac{q_\lambda q _\rho}{q^2}\right)      \right]  
%
%& a_2 = 2\nn
\nonumber
\end{align} 
These expressions are now evaluated in the following way using our ``Tensoria" technology \cite{huh13_long}: 
We first perform the trace over the Dirac indices, collecting the contracted expressions in the numerator. 
Especially for the more complicated expressions it is helpful to automate it and 
use the Feyncalc MATHEMATICA package for this \cite{feyncalc}. Then we replace the integrals of momentum written in components as described in the next section and in the Appendix of 
Ref.~\onlinecite{huh13_long}. Finally, we separate out the transverse $I^{(i)}_{\text{T}}$ 
and longitudinal $I^{(i)}_{\text{T}}$ momentum projections in the following form:
\begin{align}
\langle J^\ell_\mu(-p) J^m_\nu(p) \rangle 
&= \delta^{\ell m}\sum_{i = 0}^{2}  a_i J^{(i)} _{\mu\nu}  (p)
%\nn
\equiv 
\sum_{i=0}^2 a_i \left[ I^{(i)}_{\text{T}}(p)\left(\delta_{\mu\nu}-\frac{p_\mu p_\nu}{p^2}\right) 
+
I^{(i)}_{\text{L}}(p) \frac{p_\mu p_\nu}{p^2} 
\right]\;.
%+
%I_{\text{O}} (p) \epsilon_{\mu\nu\lambda} \bp_\lambda
\label{eq:JJ_def}
\end{align}
\subsection{Free fermion limit, $N_F\rightarrow \infty$ graph, for $C_J $}
To illustrate the procedure with a simple example, let us evaluate the leading order graph that 
also corresponds to the free fermion limit:
\begin{align}
J_{\mu\nu}^{\ell m}(p)^{(0)} &= -\text{Tr}\left[ \int_\bk \gamma_\nu T^m \frac{k_a \gamma_a }
{k^2} \gamma_\mu T^\ell \frac{(\bp+\bk)_b \gamma_b}{(\bp+\bk)^2}\right] 
\nonumber\\
&=
\delta ^{\ell m} \int \frac{ d^ 3 k}{8\pi^3}
\frac{2 k^2  \delta_{\mu\nu} + 2 \delta_{\mu\nu} k_\alpha p_\alpha - 4 k_\mu k_\nu - 2 k_\nu p_\mu - 2 k_\mu p_\nu}
{k^2 \left(\bp+\bk\right)^2}\;.
\end{align}
The integral over the first term in the numerator $2 k^2  \delta_{\mu\nu}$ is a power-law divergence in the UV and 
can be dropped.
The second, third, fourth and firth term in the numerator can be integrated using the identities
\begin{align}
\int \frac{ d^ 3 k}{8\pi^3} \frac{k_\mu}{k^2 \left(\bp+\bk\right)^2}
&=
- \frac{ p_\mu}{ 16 p }
\nonumber\\
\int \frac{ d^ 3 k}{8\pi^3} \frac{k_\mu k_\nu}{k^2 \left(\bp+\bk\right)^2}
&=
\left( 3 \frac{p_\mu p_\nu}{p^2} - \delta_{\mu \nu}\right)
\frac{p}{64}
\label{eq:twoprop}
\end{align}
with the abbreviation for the modulus $p = |\bp|$ and interchangebly $p^2 = \bp^2$. 
The result
\begin{align}
J_{\mu\nu}^{\ell m}(p)^{(0)} &= 
-\frac{p}{16}
\left(\delta_{\mu \nu} - \frac{ p_\mu p_\nu}{p^2}\right) \delta^{\ell m}
\end{align}
comes out purely transverse, 
leading to $C_J^{N_f\rightarrow \infty} = 1/16$.
Note that in order to compute $\langle T T \rangle $ (in the next section), Tensoria
performs momentum integrals of the type Eq.~(\ref{eq:twoprop}) containing up to six different momentum indices in the numerator 
and four propagators in the denominator.

\subsection{$1/N_F$ corrections to $C_J$ and discussion}
We evaluate the vertex correction and self-energy correction graphs (1) and (2) in Eq.~\ref{eq:JJ}
algorithmically and the results are in Table~\ref{tab:c_j}. 
As expected for a conserved quantity, the log-singularities of each individual graph cancel when taking 
the sum, so does the longitudinal part.
\begin{table}[t]
\centering
\begin{tabular}{ccccc}
\text{Diagram} &~ $I_T^{(i)}(p)$ ~~~& $I_L^{(i)}(p)$ & Log-Singularity (transverse) & \text{Factor $a_i$} \\[1mm]
\hline
0 & $-\frac{1}{16} p$                               & 0                                                 &       0                                                                          & 1   \\[1mm]
1 & $-\frac{p}{N_F}0.0370767 $   & $ \frac{p}{N_F}\frac{1}{3\pi^2} $ &$\frac{p}{N_F}\frac{2}{3\pi^2} \log{\frac{\Lambda}{p}}$& 1  \\[1mm]
2 & $\frac{p}{N_F}\frac{5}{36\pi^2}      $   & $-\frac{p}{N_F}\frac{1}{6\pi^2} $ &$-\frac{p}{N_F}\frac{1}{3\pi^2}   \log{\frac{\Lambda}{p}} $& 2  \\[1mm]
\hline
\end{tabular}
\caption{Evaluated contributions to the current-current correlator. 
The sum of longitudinal components off all the graphs add to 0 and the transverse parts add up to Eq.~(\ref{eq:value_cj}). 
The analytic expression for $I^{(1)}_T (p)$ (multiplied by -16) is in Eq.~(\ref{eq:JJ1})
The log-singularities mutually cancel. The self-energy correction graph (2) comes with a factor 
of $a_2=2$.}
\label{tab:c_j}
\end{table}
As announced in the Introduction, our result Eq.~(\ref{eq:value_cj}) seems to disagree with 
Chen {\it et al.}\cite{chen93} who computed $C_J$ for QED$_3$ to order $1/N_F$.
The relevant $1/N_F$ correction is given in Eq. (A17) in the appendix of their paper.
Mapping to our conventions we take $g = 1$ and $A=16$ and an overall minus sign.
These authors obtained
\begin{align}
C^{\text{Chen, et al.}}_J
=\frac{1}{16} \left( 1+ \frac{16}{N_F} \frac{3}{(2 \pi)^2} \right) 
\approx \frac{1}{16} \left( 1+ \frac{1}{N_F}1.216\right)\;,
%=\frac{1}{16} \left( 1+ \frac{1}{N} 3/(2 \pi)^2 \right) 
%\approx \frac{1}{16} \left( 1+ \frac{1}{N}0.0760\right) 
\end{align}
The sign of their $1/N_F$ correction match but the value seem to be different from Eq.~(\ref{eq:value_cj}).
As another, this time positive, cross-check, we have repeated the calculation of Appendix B in the paper by Rantner and Wen \cite{rantner02} and compared the coefficients of the logarithmically 
diverging terms in their Eq. (B10) to what we get. Both values agree to be
\begin{align}
-\frac{1}{N_F} \frac{16}{3\pi^2}  |\mathbf{p}| \ln \frac{\Lambda}{|\mathbf{p}|}\;.
\end{align}
The presence of (non-cancelling) log-singularities indicates that the quantity (staggered 
spin susceptibility in algebraic spin liquids) in their case is not conserved.
We have also computed the fermion anomalous dimension and self energy correction to order $1/N_F$ and found agreement to results from direct calculation using textbook methods (See Ref.~\onlinecite{franz02}).
\section{Tensoria technology: mini-recap}
\label{sec:tensoria}
Before proceeding, let us briefly recapitulate our algorithm 
to evaluate the tensor-valued momentum integrals as described in 
more detail in the Appendices of \cite{huh13_long,bzowski12}. 
At the heart of the algorithm are Davydychev permutation 
\cite{davydychev91,davydychev92} relations to perform integrals of the form:
\begin{align}
J_{\mu_1...\mu_M}\left(\bp_1,\bp_2,\bp_3; n; {\nu_i}\right) =
\int d^n {\bk} 
\frac{k_{\mu_1}...k_{\mu_M}}{
\left(\bk + \bp_1\right)^{2\nu_1}
\left(\bk + \bp_2\right)^{2\nu_2}
\left(\bk + \bp_3\right)^{2\nu_3}
}\;.
\label{eq:pre_davy}
\end{align}
After the Dirac traces, the integrals can all be brought into this form.
After the first momentum integration, we temporarily introduce a 
UV-momentum cutoff that formally breaks symmetries such as 
conformal invariance. Using this cutoff as a sorter, all power-law divergences 
are discarded as they would be absent in a gauge-invariant regularization schemes 
such as dimensional regularization. The remaining finite and logarithmically 
divergent terms can be integrated analytically graph-by-graph and the log-singularities 
are seen to cancel exactly.

We close this recap by noting that despite the exact cancellations of the log-singularities 
as a strong consistency check, and the many additionally performed checks of 
all sub-routines in Tensoria, at the moment we have no proof that of the exactness  
to $O(1/N_F)$ of our results for the theory Eq.~(\ref{eq:bare_action}). 
Note that the Tensoria technique was also applied in Refs.~\onlinecite{huhprl13,huh13_long}, 
for different theories. There, we found agreement with a number 
of computations using other methods.

\section{Stress energy tensor correlator $\langle T T \rangle$}
\label{sec:TT}

In this section, we extend our technology to compute the stress tensor correlator of Eq.~(\ref{eq:bare_action}) 
to next-to-leading order in $1/N_F$. We first define the stress tensor itself and write down the Feynman rules for 
the stress tensor vertices. Then, we first illustrate in some detail the calculation of the leading 
$N_F\rightarrow \infty$ graph before evaluating the remaining 8 graphs with Tensoria. The two major 
complications here are: (i) the gauge field can connect directly to the stress tensor vertex leading to a 
vertex involving 3 lines, and (ii) four 3-loop graphs, including those of the Azlamasov-Larkin type, appear. 
As in the $\langle JJ \rangle$ computation, we explicitly 
show that all log-singularities cancel when summing all graphs to ensure to 
conserved nature of $T_{\mu\nu}$ in accordance with symmetries.

\subsection{Feynman rules and graphs in momentum space}
\begin{figure}[b]
\includegraphics[width=120mm]{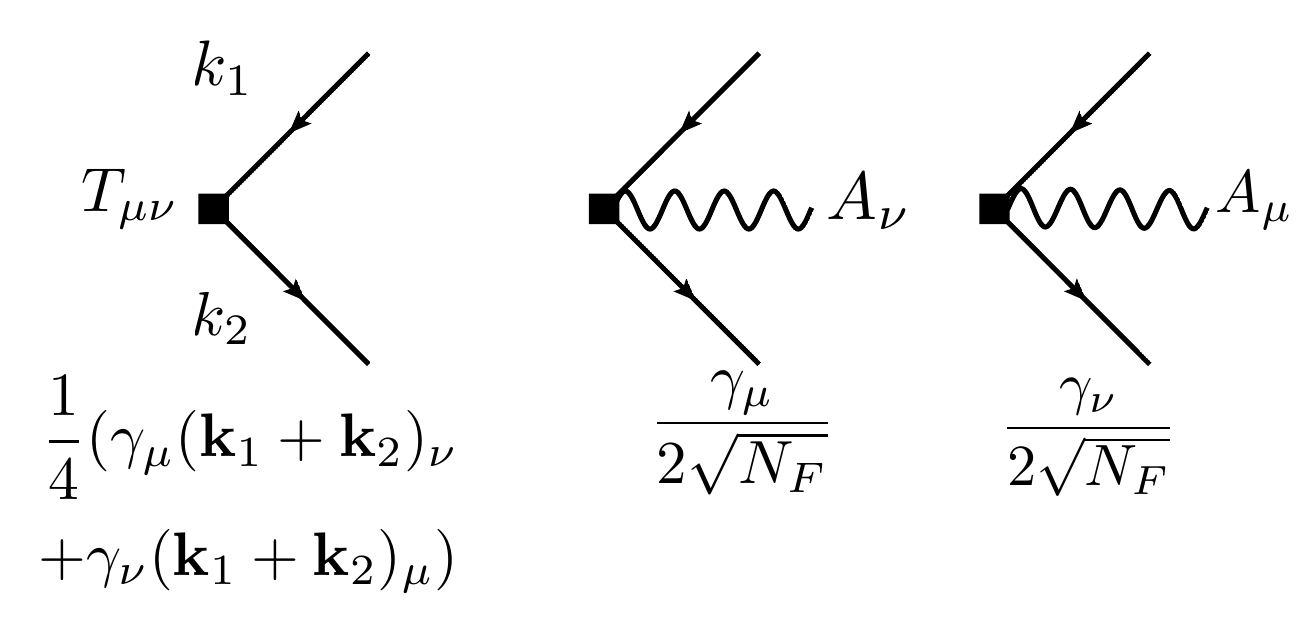}
\caption{Feyman rules for the stress tensor vertices.}
\label{fig:stressvertices}
\end{figure}

\begin{figure}[ht!]
\vspace{10mm}
\includegraphics[width=150mm]{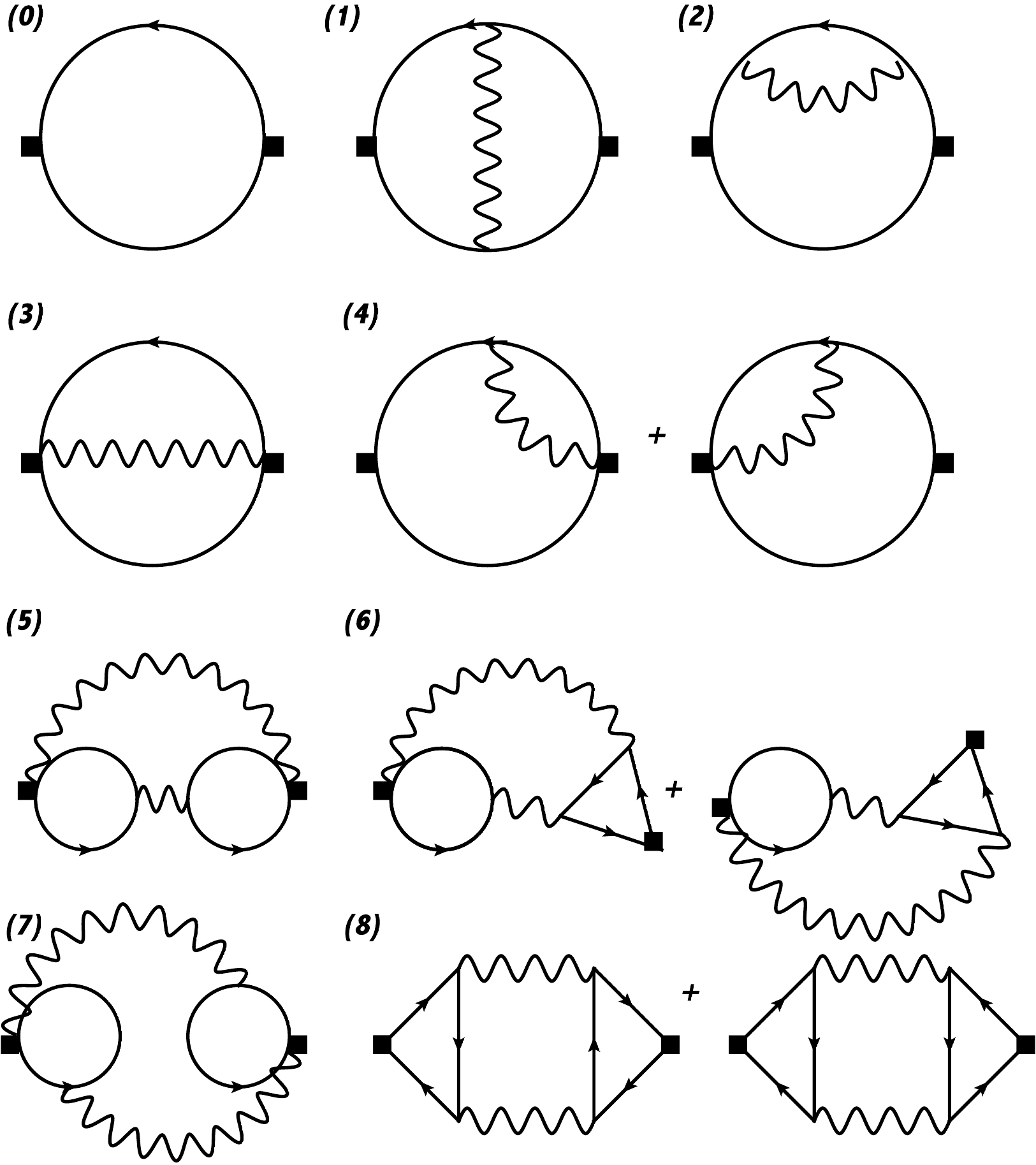}
\vspace{10mm}
\caption{Feynman diagrams contributing to the stress energy tensor correlator to next-to-leading order in $1/N_F$. Only 
diagram (0) survives in the $N_F\rightarrow \infty$ limit. Diagrams (2) and (4) come with a factor of $a_2=2$, 
$a_4=2$, respectively. The factors for the other graphs are unity $a_i=1$. The numerical values and logarithmic singularities 
for each of these graphs are exhibited in Table \ref{tab:c_t}.}
\label{fig:diagsTT}
\end{figure}

The stress tensor operator for Eq.~(\ref{eq:bare_action}) depends on both the fermions and the gauge fields 
via the gauge covariant derivative $D_\mu = \partial _\mu - iA_\mu/\sqrt{N_F}$ \cite{chowdhury13}
\begin{align} 
T_{\mu\nu} =\sum_{\alpha=1}^{N_F} \frac{i}{4} \left( \bar{\psi_\alpha} \gamma_\mu \left( D_\nu \psi_\alpha \right)  + \bar{\psi}_\alpha \gamma_\nu \left( D_\mu \psi_\alpha \right) - \left( D^\ast_\mu \bar{ \psi_\alpha} \right) \gamma_\nu \psi_\alpha - \left( D^\ast_\nu \bar{ \psi_\alpha} \right)  \gamma_\mu \psi_\alpha  \right) \;.
\end{align} 
leading to the stress tensor vertices shown in Fig.~\ref{fig:stressvertices}.
The eight graphs and their analytical expressions
shown in Figs.~\ref{fig:diagsTT}, \ref{fig:TT} contribute to order $1/N_F$ and we again denote their sum by 
\begin{align}
\langle T_{\mu\nu}(-p)T_{\lambda\rho}(p) \rangle
&= \sum_{i = 0}^{7}  a_i T^{(i)} _{\mu\nu\lambda\rho}  (p)\;.
\end{align} 
\begin{figure}[h!]
\begin{align}
T_{\mu\nu\lambda\rho}^{(0)}(p) &=  -N_F\text{Tr}\left[ \int_\bk \frac{1}{4} \gamma_\lambda (2\bk+\bp)_\rho \frac{(\bk+\bp)_a 
\gamma_a}{(\bk+\bp)^2} \frac{1}{4} \gamma_\mu (2\bk+\bp)_\nu \frac{k_b \gamma_b}{k^2}\right]  
+( \text{perm1})\nonumber\\
% &
%& a_0=1 \nn
%
T_{\mu\nu\lambda\rho}^{(1)}(p) &=  -N_F\text{Tr}\Bigg[ \int_{\bk,\bq} \frac{1}{4} \gamma_\lambda (2\bk+\bp)_\rho \frac{(\bk+\bp)_a \gamma_a}{(\bk+\bp)^2} \frac{\gamma_\epsilon}{\sqrt{N_F}} \frac{(\bk+\bp+\bq)_b \gamma_b}{(\bk+\bp+\bq)^2} \frac{1}{4} \gamma_\mu (2\bk+2\bq+\bp)_\nu \frac{(\bk+\bq)_c \gamma_c}{(\bk+\bq)^2} \frac{\gamma_\kappa}{\sqrt{N_F}} \frac{k_d \gamma_d}{k^2} 
\nonumber\\
&
\quad\quad\quad\;\;  \frac{16}{q}\left( \delta_{\kappa\epsilon} - \frac{q_\kappa q _\epsilon}{q^2}\right)      \Bigg] 
  +( \text{perm1})
  \nonumber\\
%& a_1=1 \nn
%
T_{\mu\nu\lambda\rho}^{(2)}(p) &= -N_F\text{Tr}\left[ \int_{\bk,\bq} \frac{1}{4} \gamma_\lambda (2\bk+\bp)_\rho \frac{(\bk+\bp)_a \gamma_a}{(\bk+\bp)^2} \frac{1}{4} \gamma_\mu (2\bk+\bp)_\nu \frac{k_b \gamma_b }{ k^2 } \frac{ \gamma_\epsilon }{ \sqrt{N_F} } \frac{(\bk+\bq)_c \gamma_c }{ (\bk+\bq)^2 } \frac{ \gamma_\kappa }{ \sqrt{N_F} } \frac{k_d \gamma_d }{ k^2 } 
\frac{ 16 }{ q }\left( \delta_{\kappa\epsilon} - \frac{q_\kappa q _\epsilon}{q^2}\right)  \right]  
\nonumber\\
&
\hspace{15mm}+( \text{perm1})
\nonumber\\
%& a_2 =2\nn
%
T_{\mu\nu\lambda\rho}^{(3)}(p) &= -N_F\text{Tr}\left[ \int_{\bk,\bq} \frac{ \gamma_\lambda}{2 \sqrt{N_F}} \frac{  (\bk+\bp+\bq)_a \gamma_a }{ (\bk+\bp+\bq)^2 } \frac{ \gamma_\mu}{2\sqrt{N_F}} \frac{k_b \gamma_b}{k^2} 
\frac{ 16 }{ q }\left( \delta_{\rho\nu} - \frac{q_\rho q _\nu}{q^2}\right)  \right] 
+ ( \text{perm1})
%& a_3=1 \nn
%
\nonumber\\
T_{\mu\nu\lambda\rho}^{(4)}(p) &=N_F\text{Tr}\left[ \int_{\bk,\bq}  \frac{1}{4} \gamma_\lambda (2\bk+\bp)_\rho \frac{(\bk+\bp)_a \gamma_a}{(\bk+\bp)^2}  \frac{ \gamma_\epsilon}{ \sqrt{N_F}}\frac{(\bk+\bp+\bq)_b \gamma_b}{(\bk+\bp+\bq)^2}
\frac{ \gamma_\mu}{2\sqrt{N_F}} \frac{ k_c \gamma_c }{ k^2 }
\frac{ 16 }{ q }\left( \delta_{\nu\epsilon} - \frac{q_\nu q _\epsilon }{q^2}\right)  \right] 
+( \text{perm2})
\nonumber\\
%& a_4=2 \nn
%
T_{\mu\nu\lambda\rho}^{(5)}(p) &= \int_\bq N_F\text{Tr} \left[ \int_\bk \frac{\gamma_\rho}{2 \sqrt{N_F}}  \frac{k_a \gamma_a}{k^2}  \frac{ \gamma_\epsilon }{ \sqrt{N_F} }  \frac{(\bk+\bq)_b \gamma_b }{(\bk+\bq)^2}  \right]
\frac{16}{q}  \left( \delta_{\kappa\epsilon} - \frac{q_\kappa q _\epsilon }{q^2}\right) 
\frac{16}{|\bp+\bq|}  \left( \delta_{\lambda\mu} - \frac{(\bp+\bq)_\lambda (\bp+\bq)_\mu }{(\bp+\bq)^2}\right)\times \nn
&~~~~~~
 \;\;\;N\text{Tr} \left[ \int_\bl \frac{\gamma_\nu}{2 \sqrt{N_F}} \frac{(\bl+\bq)_c \gamma_c }{(\bl+\bq)^2} \frac{ \gamma_\kappa }{ \sqrt{N_F} }   \frac{l_d \gamma_d}{l^2}   \right] 
 +(\text{perm1})
 \nonumber\\
 %& a_5=1 \nn
 %
T_{\mu\nu\lambda\rho}^{(6)}(p) &= -\int_\bq N_F\text{Tr} \left[ \int_\bk \frac{\gamma_\rho}{2 \sqrt{N_F}}  \frac{k_a \gamma_a}{k^2}  \frac{ \gamma_\kappa }{ \sqrt{N_F} }  \frac{(\bk+\bq)_b \gamma_b }{(\bk+\bq)^2}  \right]
\frac{16}{q}  \left( \delta_{\kappa\epsilon} - \frac{q_\kappa q _\epsilon }{q^2}\right) 
\frac{16}{|\bp+\bq|}  \left( \delta_{\lambda\alpha} - \frac{(\bp+\bq)_\lambda (\bp+\bq)_\alpha }{(\bp+\bq)^2}\right) \nn 
&~~~~~~~~
\Bigg(
N_F\text{Tr} \left[ \int_\bl \frac{1}{4} \gamma_\mu (\bp + 2\bl)_\nu \frac{l_c \gamma_c}{l^2} \frac{\gamma_\alpha}{\sqrt{N_F}} 
\frac{(\bp+\bq+\bl)_d \gamma_d}{(\bp+\bq+\bl)^2} \frac{\gamma_\epsilon}{\sqrt{N_F}} \frac{(\bp+\bl)_e \gamma_e}{(\bp+\bl)^2}   \right]+ \nn
&~~~~~~~~~
N_F\text{Tr} \left[ \int_\bl \frac{1}{4} \gamma_\mu (\bp+2\bq + 2\bl)_\nu \frac{(\bq+\bl)_c \gamma_c}{(\bq+\bl)^2} \frac{\gamma_\epsilon}{\sqrt{N_F}} 
\frac{l_d \gamma_d}{l^2} \frac{\gamma_\alpha}{\sqrt{N_F}} \frac{(\bp+\bq+\bl)_e \gamma_e}{(\bp+\bq+\bl)^2}   \right] 
\Bigg) + (\text{perm2})
\nonumber\\
 %& a_6=1 \nn
 %
 T_{\mu\nu\lambda\rho}^{(7)}(p) &= \int_\bq N_F\text{Tr} \left[ \int_\bk \frac{\gamma_\rho}{2 \sqrt{N_F}}  \frac{(\bk+\bp+\bq)_a \gamma_a}{(\bk+\bp+\bq)^2}  \frac{ \gamma_\epsilon }{ \sqrt{N_F} }  \frac{k_b \gamma_b }{k^2}  \right]
 \frac{16}{q}  \left( \delta_{\lambda\kappa} - \frac{q_\lambda q_\kappa }{q^2}\right) 
\frac{16}{|\bp+\bq|}  \left( \delta_{\epsilon\nu} - \frac{(\bp+\bq)_\epsilon (\bp+\bq)_\nu }{(\bp+\bq)^2}\right) \nn 
&~~~~~~~~~~
N_F\text{Tr} \left[ \int_\bl \frac{\gamma_\mu}{2 \sqrt{N_F}} \frac{(\bl+\bq)_c \gamma_c }{(\bl+\bq)^2} \frac{ \gamma_\kappa }{ \sqrt{N_F} }   \frac{l_d \gamma_d}{l^2}   \right] 
+ ( \text{perm1})
\nonumber\\
% & a_7=1 \nn
 %
 T_{\mu\nu\lambda\rho}^{(8)}(p) &= \int_\bq N_F\text{Tr} \left[ \int_\bk \frac{1}{4} \gamma_\lambda (2\bk+\bp )_\rho \frac{(\bk+\bp)_a \gamma_a}{(\bk+\bp)^2} \frac{\gamma_\alpha}{\sqrt{N_F}} 
\frac{(\bk+\bp+\bq)_b \gamma_b}{(\bk+\bp+\bq)^2} \frac{\gamma_\kappa}{\sqrt{N_F}} \frac{k_c \gamma_c}{k^2}   \right] 
\nonumber\\
&
\hspace{20mm}\frac{16}{q}  \left( \delta_{\alpha\beta} - \frac{q_\alpha q_\beta }{q^2}\right)
\frac{16}{ |\bp+\bq|}  \left( \delta_{\kappa\epsilon} - \frac{(\bp+\bq)_\kappa (\bp+\bq) _\epsilon }{(\bp+\bq)^2}\right)  \nn
&~~~~~~~~
\Bigg( N_F\text{Tr} \left[ \int_\bl \frac{1}{4} \gamma_\mu (2\bl+\bp )_\nu \frac{l_d \gamma_d}{l^2} \frac{\gamma_\beta}{\sqrt{N_F}} 
\frac{(\bl-\bq)_e \gamma_e}{(\bl-\bq)^2} \frac{\gamma_\epsilon}{\sqrt{N_F}} \frac{(\bl+\bp)_f \gamma_f}{(\bl+\bp)^2}   \right]+ \nn
&~~~~~~~~~~
N_F\text{Tr} \left[ \int_\bl \frac{1}{4} \gamma_\mu (2\bl+\bp )_\nu \frac{l_d \gamma_d}{l^2} \frac{\gamma_\epsilon}{\sqrt{N_F}} 
\frac{(\bl+\bp+\bq)_e \gamma_e}{(\bl+\bp+\bq)^2} \frac{\gamma_\beta}{\sqrt{N_F}} \frac{(\bl+\bp)_f \gamma_f}{(\bl+\bp)^2}   \right] 
\Bigg) 
+ ( \text{perm1})
\nonumber
% & a_8=1 \nn
%\label{eq:TT}
\end{align} 
\caption{Analytical expressions for the 8 graphs in Fig.~\ref{fig:diagsTT}.
Here, ``perm1'' indicate permutations $(\mu \leftrightarrow \nu)$, $(\lambda \leftrightarrow \rho)$, and  $(\mu \leftrightarrow \nu,  \lambda \leftrightarrow \rho )$ and ``perm2'' indicate $(\mu \leftrightarrow \lambda,\nu\leftrightarrow \rho)$, switched as a pair, in addition to permutations indicated by ``perm1''. ``perm2'' will increase the number of terms by a factor of 8. }
\label{fig:TT}
\end{figure}

\newpage
In order to compute the ``central charge'' $C_T$, we will project it out from the evaluated graphs using the relation 
Eq.~(\ref{eq:TTgeneral}):
\begin{align}
C_T = \frac{1}{4 |\bp|^3} \delta_{\mu\lambda}\delta_{\nu\rho}\langle T_{\mu\nu}(-p)T_{\lambda\rho}(p) \rangle\;.
\end{align}
 We note here that a number of previous analyses \cite{petkou94,petkou95,petkou96} have been conducted in real space, 
 where the invariance of correlators under the full set of conformal transformations are transparent but the analysis to work 
 out the constants for an interacting CFT is quite involved.
 
 \subsection{Free fermion limit, $N_F\rightarrow \infty$ graph, for $C_T $}
 Let us evaluate the leading order graph, the first line in Fig.~\ref{fig:TT} that 
 also corresponds to the free fermion limit.
 Including the index permutations described in the caption of the figure, we have
\begin{align}
 \frac{C_T^{N_F \rightarrow \infty}}{N_F}
 &=
 \frac{1}{4 |\bp|^3} \delta_{\mu\lambda}\delta_{\nu\rho} T^{(0)}_{\mu\nu \lambda\rho}(p)
 \nonumber\\
 &=  \frac{1}{4 |\bp|^3} \frac{1}{2}\int \frac{ d^3 k}{8\pi^3}
 \frac{ k_\alpha p_\alpha k_\beta p_\beta - k^2 p^2}
 {k^2 \left( \bk + \bp\right)^2}
 \nonumber\\
 & =
  \frac{1}{4 |\bp|^3} 
 \frac{1}{2} p_\alpha p_\beta 
 \left( 3 \frac{p_\alpha p_\beta}{p^2} - \delta_{\alpha \beta}\right)
\frac{p}{64}
\nonumber\\
&=
\frac{1}{256}\,
\end{align}
where we dropped the second term in the numerator in the second line because 
it is a power-law divergence in the UV, absent in dimensional regularization. We can also check that without immediately 
contracting the graph, the uncontracted terms 
fulfill the index structure of Eq.~(\ref{eq:TTgeneral}).

\subsection{$1/N_F$ corrections for $C_T$ and discussion }
Tensoria computes the $1/N_F$ corrections algorithmically and Table~\ref{tab:c_t} collects the results.
\begin{table}[ht!]
\centering
\begin{tabular}{ccccc}
\text{Diagram} &~ $C_T^{(i)}$ ~~~ & Log-Singularity& \text{Factor $a_i$} \\[1mm]
\hline
0 &   $\frac{N_F}{256}$                           & 0                                                                              & 1   \\[1mm]
1 &   	$-0.00162$				     & $-\frac{7}{120 \pi^2}p^3\log{\frac{\Lambda}{p}} $                   	& 1  \\[1mm]
2 &   	$\frac{1}{576 \pi^2}$		              & $\frac{1}{48 \pi^2}p^3\log{\frac{\Lambda}{p}} $ 			& 2  \\[1mm]
3 &   	$0$				                      & $0 $  			& 1  \\[1mm]
4 &   	$-\frac{19}{288 \pi^2}$		     & $-\frac{1}{24 \pi^2}p^3\log{\frac{\Lambda}{p}} $ 			& 2  \\[1mm]
5 &   	$0$        				              & $0 $ 			& 1  \\[1mm]
6 &   	$-\frac{1}{128}+\frac{19}{144 \pi^2}$				& $\frac{1}{12 \pi^2}p^3\log{\frac{\Lambda}{p}} $  			& 1  \\[1mm]
7 &   $\frac{1}{256}$				& $0  $  			& 1  \\[1mm]
8 &   	$\frac{1}{256}+\frac{17}{720\pi^2}$				& $\frac{1}{60 \pi^2}p^3\log{\frac{\Lambda}{p}}  $  			& 1  \\[1mm]
\hline
\end{tabular}\\[5mm]
\caption{Evaluated contributions to the stress tensor correlator and the log-singularities. 
The log-singularities cancel exactly after summing all graphs. The analytic expression for $C_T^{(1)}$ (multiplied by 256) 
is given in Eq.~(\ref{eq:ct1}).}
\label{tab:c_t}
\end{table}
As before, we observe an exact cancellation 
of the logarithmic singularities of each graph in accordance with symmetry 
requirements. Summing the graphs leads to Eq.~(\ref{eq:CT_final}) in the Introduction. 
Note that the contributions of $1/N_F$ correction graphs 1 - 8 to $C_T$ 
do not have definite sign: 1 and 4 are negative while the others are positive.
As already touched upon in the Introduction, it will be interesting to understand 
the signs and structure of the interaction corrections to $C_T$ for more general 
IR and UV fixed points especially against the backdrop of ``$C_T$ measuring the number of
degrees of freedom'' of a given field theory.

%{\color{red} 
%In addition to the discussion 
%in the Introduction, we mention here that the components of the stress tensor correlator 
%also yield the shear viscosity particularly relevant for strongly interacting 
%quantum field theories at finite temperature \cite{kss05,enss11}. 
%In order to resolve the collisional physics, 
%however, it is necessary to solve a Boltzmann equation or invoke the AdS-CFT 
%correspondence (see e.g.: Refs.~\onlinecite{enss11,katz14} 
%and references therein).}

\section{Conclusions}
The aim of this paper was to provide precision computations of the ``central charge''
$C_T$ and universal conductivity $C_J$ of interacting conformal field theories 
in $2+1$ dimensions. We considered $N_F$ Dirac fermions coupled 
to an ``emergent photon'' motivated by frequent occurrence of this field theory 
in a variety of condensed matter systems. The low-energy sector is also equivalent 
to many-flavor QED$_3$ in the conformal phase.

Our hope is that our results could become a useful diagnostic for numerical 
evaluations of entanglement properties of CFT$_3$'s, conformal bootstrap approaches, 
or application of the AdS-CFT correspondence.
Going forward, our technology may also complement explicit computations of conformal 
correlators in the context of dualities of Large $N$ Chern-Simons Matter 
Theories \cite{aharony12, gur-ari13,aharony13}. In particular, one may be able to 
directly compute higher-order current and stress tensor correlators from 
``both sides of the duality", taking for example fermionic matter fields coupled $U(N_F)_{k_F}$ Chern-Simons 
on one side and the $U(N_b)_{k_b}$ critical-bosonic Chern-Simons vector model on the other side,
and checking the parameter space for the conjectured duality.

\acknowledgments

We thank Andrea Allais, Holger Gies, Zohar Komargodski, Jan M. Pawlowski, and Silviu Pufu for 
discussions and Subir Sachdev for guidance, collaboration on related projects, 
and critically reading the manuscript. We also thank Simone Giombi, Grigory 
Tarnopolsky and Igor Klebanov for a correspondence 
that led to the clarification of the sign error in previous 
versions of the paper. This research was supported by the 
Leibniz prize of A. Rosch, and the NSF grant DMR-1360789. This research was also supported 
in part by Perimeter Institute for Theoretical Physics. Research at Perimeter Institute is 
supported by the Government of Canada through Industry Canada and by the Province of Ontario 
through the Ministry of Economic Development \& Innovation.


\begin{thebibliography}{}

\bibitem{wilsonfisher71}
K. G. Wilson, and M. E. Fisher,
{\it Critical Exponents in 3.99 Dimensions},
Phys. Rev. Lett. {\bf 28}, 240 (1971).

\bibitem{abe73}
R. Abe,
{\it Critical exponent $\eta$ up to $1/N^2$ 
for the Three-Dimensional System with Short-Range Interaction},
Prog. of Theor. Phys., {\bf 49}, 6 (1973).

\bibitem{petkou95}
A. Petkou,
{\it $C_T$ and $C_J$ up to next-to-leading order in 
$1/N$ in the conformally invariant $O(N)$ vector 
model for $2 < d < 4$},
Phys. Lett. B {\bf 359}, 101 (1995).

\bibitem{showk12}
S. El-Showk, M. F. Paulos, D. Poland, S. Rychkov, D. Simmons-Duffins,
and A. Vichi,
{\it Solving the 3D Ising model with the conformal bootstrap},
Phys. Rev. D {\bf 86}, 025022 (2012).

\bibitem{showk14}
S. El-Showk, M. F. Paulos, D. Poland, S. Rychkov, D. Simmons-Duffins,
and A. Vichi,
{\it Solving the 3D Ising model with the conformal bootstrap II. c-Minimization 
and Precise Critical Exponents},
arXiv:1403.4545 (2014).

\bibitem{cha91}
M. C. Cha, M. P. A. Fisher, S. M. Girvin, M. Wallin,
and A. P. Young,
{\it Universal conductivity of two-dimensional films 
at the superconductor-insulator transition},
Phys. Rev. B {\bf 44}, 6883 (1991).

\bibitem{fazio96}
R. Fazio and D. Zappala,
{\it $\epsilon$ expansion of the conductivity at the 
superconductor-Mott-insulator transitions},
Phys. Rev. B {\bf 53}, R8885 (1996).

\bibitem{nelson89}
S. Chakravarty, B. I. Halperin, and 
D. R. Nelson, 
{\it Two-dimensional quantum Heisenberg antiferromagnet at low 
temperatures},
Phys. Rev. B {\bf 39}, 2344 (1994).

\bibitem{chubukov94}
A. V. Chubukov, S. Sachdev,
{\it Theory of two-dimensional quantum Heisenberg 
antiferromagnet with a nearly critical ground state},
Phys. Rev. B {\bf 49}, 11919 (1994).

\bibitem{kaul08}
R. K. Kaul, and S. Sachdev,
{\it Quantum criticality of $U(1)$ gauge theories 
with fermionic and bosonic matter in two spatial dimensions},
Phys. Rev. B {\bf 77}, 155105 (2008).

\bibitem{chen92}
W. Chen, G. W. Semenoff, and Y.-S. Wu,
{\it Two-loop analysis of non-Abelian Chern-Simons theory},
Phys. Rev. D {\bf 46}, 5521 (1992).

\bibitem{chen93} W. Chen, M.~P.~A.~Fisher, and Y.-S. Wu, 
{\it Mott transition in an anyon gas},
Phys. Rev. B {\bf 48}, 13749 (1993).

\bibitem{senthil04}
T. Senthil, A. Vishwanath, L. Balents, S. Sachdev, 
and M. P. A. Fisher,
{\it Deconfined Quantum Criticality},
Science {\bf 303}, 1490 (2004).

\bibitem{sandvik07}
A. W. Sandvik,
{\it Evidence for deconfined quantum criticality in a 
two-dimensional Heisenberg model with four-spin 
interactions},
Phys. Rev. Lett. {\bf 98}, 227202 (2007).

\bibitem{huhprl13}
Y. Huh, P. Strack, and S. Sachdev,
{\it Vector Boson Excitations Near Deconfined Quantum Critical Points},
Phys. Rev. Lett. {\bf 111}, 166401 (2013).

\bibitem{rantner02}
W. Rantner, and X.-G. Wen,
{\it Spin correlations in the algebraic spin liquid:
Implications for high-T$_c$ superconductors},
Phys. Rev. B {\bf 66}, 144501 (2002).

\bibitem{franz02}
M. Franz, Z. Tesanovic, and O. Vafek,
{\it QED$_3$ theory of pairing pseudogap in cuprates: 
From d-wave superconductor to antiferromagnet via an 
algebraic Fermi liquid},
Phys. Rev. B {\bf 66}, 054535 (2002).

\bibitem{franz03}
M. Franz, T. Pereg-Barnea, D. E. Sheehy, and Z. Tesanovic,
{\it Gauge-invariant response functions in algebraic Fermi liquids},
Phys. Rev. B {\bf 68}, 024508 (2003).

\bibitem{kaul07}
R. K. Kaul, Y.-B. Kaim, S. Sachdev, and T. Senthil,
{\it Algebraic charge liquids},
Nature Physics {\bf 4}, 28 (2007).

\bibitem{cardy08}
J. Cardy,
{\it Conformal Field Theory and Statistical Mechanics},
arXiv:0807.3472 (2008).

\bibitem{polyakov87}
A. M. Polyakov, 
{\it Gauge Fields and Strings} (Harwood Academic, Chur, 1987).

\bibitem{coleman88}
S. Coleman, {\it Aspects of Symmetry}
(Cambridge University Press, Cambridge, UK, 1988).

\bibitem{appelquist86}
T. W. Appelquist, M. Bowick, D. Karabali, and L. C. R. Wijewardhana,
{\it Spontaneous chiral-symmetry breaking in three-dimensional QED},
Phys. Rev. D {\bf 33}, 3704 (1986).

\bibitem{appelquist88}
T. Appelquist, D. Nash, and L.C.R. Wijewardhana,
{\it Critical Behavior in (2+1)-Dimensional QED},
Phys. Rev. Lett. {\bf 60}, 2575 (1988).

\bibitem{nash89}
D. Nash,
{\it High-Order Corrections in (2+1)-Dimensional QED},
Phys. Rev. Lett. {\bf 62}, 3024 (1989).

\bibitem{son07}
D. T. Son,
{\it Quantum critical point in graphene approached 
in the limit of infinitely strong Coulomb interaction},
Phys. Rev. B {\bf 75}, 235423 (2007).

\bibitem{juricic10}
V. Juricic, O. Vafek, and I. F. Herbut,
{\it Conductivity of interacting massless Dirac particles 
in graphene: Collisionless regime},
Phys. Rev. B {\bf 82}, 235402 (2010).

\bibitem{herbut13}
I. F. Herbut, and V. Mastropietro,
{\it Universal conductivity of graphene 
in the ultrarelativistic regime},
Phys. Rev. B {\bf 87}, 205445 (2013).

\bibitem{kotikov14}
A. V. Kotikov and S. Teber,
{\it Two-loop fermion self-energy in 
reduced quantum electrodynamics and 
application to the ultrarelativistic limit of 
graphene},
Phys. Rev. D {\bf 89}, 065038 (2014).

\bibitem{barnes14}
E. Barnes, E. H. Hwang, R. E. Thockmorton, 
and S. Das Sarma,
{\it Effective field theory, three-loop perturbative 
expansion, and their experimental implications in graphene
many-body effects},
Phys. Rev. B {\bf 89}, 235431 (2014).

\bibitem{braun14}
J. Braun, H. Gies, L. Janssen, D. Roscher,
{\it Phase structure of many-flavor QED$_3$},
Phys. Rev. D {\bf 90}, 036002 (2014).

\bibitem{huh13_long}
Y. Huh, P. Strack, and S. Sachdev, 
{\it Conserved current correlators of conformal field theories 
in $2+1$ dimensions},
Phys. Rev. B {\bf 88}, 155109 (2013).

\bibitem{cardy10}
J. Cardy,
{\it The ubiquitous 'c': from the Stefan-Boltzmann Law 
to Quantum Information}, J. Stat. Mech. 1010:P10004 (2010).

\bibitem{perlmutter14}
E. Perlmutter,
{\it A universal feature of CFT R\`enyi entropy},
JHEP {\bf 03}, 117 (2014).

\bibitem{kaul13}
R. K. Kaul, R. G. Melko, and A. W. Sandvik,
{\it Bridging Lattice-Scale Physics 
and Continuum Field Theory with Quantum 
Monte Carlo Simulations},
Annu. Rev. Condens. Matter Phys. {\bf 4}, 179 (2013).

\bibitem{hathrell_scalar}
S. J. Hathrell, 
{\it Trace Anomalies and $\lambda \phi^4$ Theory in Curved Space},
Annals of Physics {\bf 139}, 136 (1982).

\bibitem{hathrell_qed}
S. J. Hathrell,
{\it Trace Anomalies and QED in Curved Space},
Annals of Physics {\bf 142}, 34 (1982).

\bibitem{jack84}
I. Jack and H. Osborn,
{\it Background field calculations in curved spacetime:
(I). General application and application to scalar fields},
Nuclear Physics B {\bf 234}, 331 (1984).

\bibitem{jack85}
I. Jack,
{\it Background field calculations in curved spacetime:
(III). Application to a general gauge theory 
to fermions and scalars},
Nucl. Phys. B {\bf 253}, 323 (1985).

\bibitem{cappelli91}
A. Cappelli, D. Friedan, and J. I. LaTorre,
{\it c-Theorem and spectral representation},
Nucl. Phys. B {\bf 352}, 616 (1991).

\bibitem{petkou94}
H. Osborn, and A. Petkou, 
{\it Implications of Conformal Invariance in Field Theories 
for General Dimensions},
Ann. Phys. {\bf 231}, 311 (1994).

\bibitem{petkou96}
A. Petkou,
{\it Conserved Currents, Consistency Relations, and 
Operator Product Expansions in the Conformally Invariant 
$O(N)$ Vector Model},
 Ann. of Phys. {\bf 249}, 180 (1996).
 
\bibitem{zoller12}
M. F. Zoller and K. G. Chetyrkin
{\it OPE of the energy-momentum tensor correlator 
in massless QCD},
JHEP {\bf 12}, 119 (2012).

\bibitem{chowdhury13}
D. Chowdhury, S. Raju, S. Sachdev, A. Singh, and P. Strack,
{\it Multipoint correlators of conformal field theories: 
Implications for quantum critical transport},
Phys. Rev. B {\bf 87}, 085138 (2013).

\bibitem{maldacena11}
J.M. Maldacena and G.L. Pimentel,
{\it On graviton non-gaussianities during inflation},
JHEP {\bf 09}, 045 (2011).

\bibitem{klebanov12}
I. R. Klebanov, S. S. Pufu, S. Sachdev, and B. R. Safdi,
{\it R\'enyi entropies for free field theories},
JHEP {\bf 04}, 074 (2012).

\bibitem{dymarsky13}
A. Dymarsky, Z. Komargodski, A. Schwimmer, and S. Theisen,
{\it On Scale and Conformal Invariance in Four Dimensions},
arXiv:1309.2921 (2013).

\bibitem{bzowski14}
A. Bzowski and K. Skenderis,
{\it Comments on scale and conformal invariance 
in four dimensions},
arXiv:1402.3208 (2014).

\bibitem{myers11}
R. C. Myers and A. Sinha,
{\it Holographic c-theorems in arbitrary dimension},
JHEP {\bf 01}, 125 (2011).

\bibitem{klebanov11}
I. R. Klebanov, S. S. Pufu, and B. R. Safdi,
{\it F-theorem without supersymmetry},
JHEP {\bf 10}, 038 (2011).

\bibitem{appelquist99}
T. Appelquist, A. G. Cohen, and M. Schmaltz,
{\it A new constraint on strongly coupled field theories},
Phys. Rev. D {\bf 60}, 045003 (1999).

\bibitem{cardy87}{J. Cardy, 
{\it Anisotropic corrections to correlation functions in finite-size 
systems}, Nucl. Phys. B {\bf 290}, 355 (1987).}

\bibitem{feyncalc}
{\it Tools and Tables for Quantum Field Theory Calculations},
URL: http://www.feyncalc.org/.

\bibitem{bzowski12}
A. Bzowski, P. McFadden, and K. Skenderis,
{\it Holographic predictions for cosmological 3-point functions},
JHEP {\bf 03}, 091 (2012).

\bibitem{davydychev91}
A.I. Davydychev,
{\it A simple formula for reducing Feynman diagrams to scalar integrals},
Phys. Lett. B {\bf 263}, 107 (1991).

\bibitem{davydychev92}
A.I. Davydychev,
{\it Recursive algorithm for evaluating vertex-type Feynman integrals},
J. Phys. A: Math. Gen. {\bf 25}, 5587 (1992).

%\bibitem{kss05}
%P. K. Kovtun, D. T. Son, and A. O. Starinets,
%{\it Viscosity in Strongly Interacting Quantum Field Theories 
%from Black Hole Physics},
%Phys. Rev. Lett. {\bf 94}, 111601 (2005).

%\bibitem{enss11}
%{T. Enss, R. Haussmann, and W. Zwerger},
%{\it Viscosity and scale invariance in the unitary Fermi gas},
%Ann. of Phys. {\bf 326}, 770 (2011).

%\bibitem{katz14}
%E. Katz, S. Sachdev, E. S. Sorensen, and W. Witczak-Krempa,
%{\it Conformal field theories at finite temperature: 
%operator product expansions, Monte Carlo, and holography},
%arXiv:1409:3841 (2014).

\bibitem{aharony12}
O. Aharony, G. Gur-Ari, and R. Yacoby,
{\it Correlation Functions of Large N Chern-Simons-Matter 
Theories and Bosonization in Three Dimensions},
JHEP {\bf 12}, 028 (2012). 

\bibitem{gur-ari13}
G. Gur-Ari and R. Yacoby
{\it Correlators of large $N$ fermionic Chern-Simons 
vector models},
JHEP {\bf 02}, 150 (2013).

\bibitem{aharony13}
O. Aharony, S. Giombi, G. Gur-Ari, J. Maldacena, and R. Yacoby,
{\it The thermal free energy in large $N$ 
Chern-Simons-matter theories},
JHEP {\bf03}, 121 (2013).


\end{thebibliography}
\end{document}